\documentclass[12pt]{iopart}
 \usepackage{graphicx}
\usepackage{float}

\begin{document}

\title {Introduction to {\it Quantum Matter}}
\author{Frank Wilczek}
\address{Center for Theoretical Physics and Department of Physics,
 Massachusetts Institute of Technology, Cambridge, MA 02139}
 \ead{wilczek@mit.edu}
 
 \begin{abstract}
This paper records my introductory remarks at Nobel Symposium 148, on Graphene and Quantum Matter, at Saltsj\"obaden, Sweden, in June 2010.   After some broad comments on the quantum theory of matter as a frontier of physics, and some slightly more particular comments about re-quantization, I report on the universal geometry that arises in a refined discussion of quantum-mechanical level crossing. 
 \end{abstract}
 
 
\maketitle

%

The quantum theory of matter is still young.   It's been over 100 years since its birth, in the work of Max Planck; very close to 100 years since Bohr's atomic model, which made it plausible that quantum theory held the key to understanding the deep structure of matter; and about 85 years since the foundational ideas of quantum mechanics were proposed by Heisenberg, Schr\"odinger, and other greats.  While those ideas have been augmented with many new ones, and fruitfully applied in many directions, their essence remains unchanged.   Yet such is their depth, power, and strangeness that they continue to conjure up surprises, and to tantalize us with their potential.   

This {\it Nobel Symposium\/} is dedicated to quantum matter in general, with special emphasis on a spectacular success story of recent years, the discovery and exploration of graphene\footnote{The word ``discovery'' suggests a discontinuity, whereas science is a complex, cumulative process.   What should be uncontroversial is that the work of Geim, Novoselov, and their collaborators rapidly brought the subject to a new level of scientific interest.   A few months after the Symposium, Geim and Novoselov were awarded the Nobel Prize for their work.}.     

This introductory talk has three parts.   I'll start with a few general reflections on the quantum theory of matter, emphasizing some recently emerging themes.   Then I'll briefly introduce a remarkable class of developments that are close to my heart: The embodiment, in materials, of concepts -- and even equations -- that were studied earlier for their possible application to fundamental physics, or simply for their intrinsic beauty.   Finally, to exhibit one distilled essence of that theme, I'll go into some detail about a basic problem in elementary quantum mechanics, the problem of level crossing, and show how vortices, monopoles, and even $SU(2)$ instantons emerge in that context.   

\section{Quantum Phenomenology and Quantum Engineering}

The first task of quantum theory, as of all scientific theories, is to describe the natural world.   The most basic ideas of quantum theory, including wave functions, entanglement (implicit already in having wave functions in configuration space), superposition, commutation relations, the Schr\"odinger equation, quantized angular momentum and spin, and Bose-Einstein and Fermi-Dirac statistics, emerged from confrontations with concrete phenomena of radiation and atomic spectroscopy.   Fundamental insights and techniques for deploying those concepts, notably including the application of symmetry arguments and group theory, emerged from engagement with more intricate facts of spectroscopy (including molecular and nuclear spectroscopy), chemistry, and the theory of solids.   

Phenomenology continues to pose new challenges for the quantum theory of matter.   Many specific issues remain outstanding, including the problem of strange metals and high-temperature superconductivity, and puzzling marvels from the biological world, notably photosynthesis.   There's also a broader and more general challenge.  There is every reason to think that for any practical question of chemistry and materials science we can write down appropriate equations, whose solution should provide the answer.   In that precise and powerful -- if {\it ultimately\/} limited -- sense, the ``Theory of Everything'' is in our grasp.   The challenge, of course, is to solve the equations in useful detail, in as many contexts as possible.   

After the natural world, a second frontier beckons.    We needn't be content with the world as we find it.  We can build on our insights to imagine, and construct, materials and devices that Nature doesn't provide.   

Semiconductor and microelectronic technology, laser technology, and molecular design would be inconceivable without modern quantum theory.   It's fair to call them examples of quantum engineering.      Modern quantum theory underlies the rapid rise of graphene, and also fullerenes and nanotubes, as well.  Major properties of that material including its strength, its remarkable band structure (and that of its bilayer variant), its signature quantum Hall effect, and others are direct and fairly straightforward consequences of basic quantum theory; indeed they were largely anticipated in advance of graphene's experimental discovery.      

Quantum engineering, clearly, is already a diverse and highly successful enterprise.   But it is still in its infancy.   The Hilbert space of many-body systems is enormously large, and at present we can only reach small corners of it, with limited navigational control.   A great task for the future will be to embody the rich possibilities of entanglement within systems that we can control, and that are not cripplingly fragile.   Quantum simulators, or ultimately general purpose quantum computers, might enable us to enter a virtuous cycle, whereby better solutions to the equations of quantum theory lead to better quantum machines lead to better solutions of the equations ... .   At that point, quantum theory will have entered its adolescence.

\section{Artificial Atoms and Ethers}

Well before the emergence of modern ideas about atomic structure, in the nineteenth century, Thomson and Tait \cite{thomsonTait}, impressed with the stability of hydrodynamic vortices, proposed that atoms are vortices, or knots of vortices, in the ether.  They produced models of such vortex-atoms experimentally, in the form of smoke rings.   Tait, especially, elaborated the idea that different elemental atoms correspond to topologically distinct knotted vortices; his detailed tabulation of knots, in pursuit of that idea, inaugurated a still-flourishing field of mathematics.   

As a theory of natural atoms those ideas were, we now know, wide of the mark.   But it's easy to understand their appeal: Vortices, even in the form of smoke-rings, can have an impressive degree of stability; they can be knotted into topologically distinct forms, which are also quasi-stable; and their interactions are complex and intricate, yet reproducible.     Those would be attractive features to embody in artificial ``atoms'', meant to be building-blocks for quantum engineering.   

\subsection{Designer Atoms and Re-quantization}

Nor is that possibility entirely fanciful.  Vortices in type II superconductors dominate their response to magnetic fields.   They are quite tangible objects that experimenters work with, manipulate, and even image.   Several other quasi-macroscopic objects, including notably small magnetic domains, or the walls that separate them, share some of the same characteristics.   The ``artificial ethers'' in which these objects live are also subjects for creative engineering.   They can, for example, be effectively two-dimensional, or even one-dimensional, if we work below the energy gap for motion in the transverse directions.  

The study of artificial atoms inside artificial ethers has already brought us many interesting surprises.  Particularly interesting, I think, is the phenomenon of {\it re-quantization}.    That is: The quantum numbers of the elementary excitations inside artificial ethers can be  different from the quantum numbers of any elementary quanta we find in the vacuum.   In particular, we can find fractional electric charge, electron number, or angular momentum.    The root of these phenomena is mixing between ordinary and topological quantum numbers\footnote{This viewpoint, elaborated below, was inspired by conversations with Sidney Coleman.}.    

In quantum electrodynamics we learn that there is a universal factor relating bare charges, that occur in the formulation of the theory as coefficients parameterizing interactions of quantum fields, to the observed (renormalized) charges of particles.   Technically, this appears as a consequence of Ward's identity.    A heuristic explanation is that dielectric response of long-wavelength virtual particles, which governs the renormalization of charge, can only be governed by structure that extends to spatial infinity, {\it viz}. the value of the conserved, integrated charge.   

If, however, there is additional structure -- {\it e.g}., two distinct conserved quantities -- then vacuum polarization can mix things up.  Consider, to be concrete, artificial atoms in artificial ethers that have conserved topological, as well ordinary electric, charges.   Then the observed charge spectrum, in terms of renormalized charges, be influenced by both.   Since the renormalizing effects due to several atoms are additive, the observed charge will take the form
\begin{equation}\label{reQuantization}
Q/e ~=~  q + \frac{1}{\pi} \arctan r \ \tau ~=~ q + \frac{\theta}{2\pi} \tau 
\end{equation}
(The equivalent forms on the right-hand side are natural in different contexts \cite{GW} \cite{witten}.)
Here $q$, representing the fundamental quantum numbers created by elementary fields, is an integer.   But in the presence of topological invariants, typically parameterized by discrete quantities $\tau$, the observed charges $Q/e$ will not be integers.  The charge spectrum, though modified, remains highly structured.  We might say charge has been {\it re}-quantized, as opposed to {\it de}-quantized.

In the first form, it is convenient to introduce $\phi \equiv \arctan \, r$.   
Here the notations $\phi, \theta$ suggest angles; since an integer can be absorbed into a shift in $q$, the physics of \eref{reQuantization} is periodic in $\phi \rightarrow \phi +  \pi$ or $\theta \rightarrow \theta + 2\pi$.   In different physical examples, several cases have arisen: 
\begin{itemize}
\item If the topological quantum number $\tau$ is itself an additive integer, generally $\theta$ is simply a number, characteristic of the material and disturbances in question.    
\item If the topological quantum number is an integer modulo $m$, then we must have
\begin{equation}
\theta ~=~ \frac{2\pi k}{m}
\end{equation}
with charges
\begin{equation}
Q/e ~=~ q + \frac{k}{m} \tau
\end{equation}
for some integer $k$; this allows $m$ of our atoms to annihilate yielding states with conventional charges.  This situation arises in the fractional quantum Hall effect.
\item In particular, if our atoms can annihilate in pairs, so $m=2$, then the only non-trivial possibility is $k=1$, implying half-integer charges for $\tau = 1 \equiv -1$.   This situation is typical of self-conjugate ({\it e.g}. $C$ invariant) atoms.   Symmetry demands that the minimal charges $\pm \frac{1}{2}$ must be realized by degenerate states; therefore we must have zero-energy modes connecting them.   Historically this was the first case of re-quantization to be analyzed, by Jackiw and Rebbi \cite{jr}. 
\end{itemize}

\subsection{Designer World-Lines}
 
Still more interesting and open-ended possibilities arise when we extend consideration of artificial ethers in two dimensions, to take the dimension of time into account.  For then we are dealing with three-dimensional space-time; and in those three dimensions we can imagine the world-lines of artificial atoms tangled into knots.   That thought brings us, of course, into the world of anyons (as the preceding discussion alluded to the world of topological solitons).

\section{Geometry of Level Crossing}

Frequently in quantum mechanics we are led to consider families of Hamiltonians $H(\lambda^j)$ that depend on a parameter.  For example, in the context of the Born-Oppenheimer approximation the $\lambda^j$ might be nuclear positions; or in band theory they might be labels for the band numbers, and quasi-momenta.     For each value of $\lambda^j$ the energy eigenspaces are simply linear manifolds, but when we consider how families of such spaces fit together non-trivial geometry comes into play.  

In these and other applications, level crossings play an important role.   As we'll see, some canonical geometric and topological structures characterize quantum-mechanical level crossings.  Those structures are at the root of several phenomena of current interest, including the quantum Hall effect in general, the anomalous quantum Hall effect in graphene, quantized transport in general, and, specifically, quantized transport in topological insulators.   

\subsection{The Original Case}

The simplest case of level crossing was analyzed already in the earliest days of quantum theory, by von Neumann and Wigner in 1927 \cite{vNWigner}.  It was only many years later, in 1963, that Herzberg and Longuet-Higgins \cite{HerzbergL-H} pointed out an additional subtlety, that is the real beginning of our story. 

In this simplest case we focus on two nearby but otherwise isolated energy levels, and suppose that the relevant Hamiltonians are real (and, of course, Hermitean).  Then the eigenstates can also be taken to be real.   Shedding some trivial complications, by adding $\lambda^j$-dependent constants to the Hamiltonians $H(\lambda^j)$, we can suppose that the crossing occurs at zero energy, and that the average energy is zero.   Then near the crossing, to linear order -- and of course neglecting any other levels -- the possible Hamiltonians  their associated energies can be parameterized as
\begin{eqnarray}
H(x, y) ~&=&~ \left(\begin{array}{cc}x & y \\y & -x\end{array}\right) \\
~&=&~ y \sigma_1 + x \sigma_3 \label{sigmaForm} \\
~&=&~ \sqrt{x^2 + y^2} e^{-i\frac{\phi}{2}\sigma_2}\sigma_3e^{i\frac{\phi}{2}\sigma_2} \label{diagonalization} \\
\phi ~&\equiv&~ \tan^{-1} \frac{y}{x} 
\end{eqnarray}
with energy eigenvalues
\begin{equation}
E_{\pm} ~=~ \pm \sqrt{x^2 + y^2}
\end{equation}
Here $x$ and $y$ are real numbers.  From this two simple but important consequences follow:
\begin{enumerate}
\item If only one parameter is in play, generically crossing is avoided. Crossing is a robust phenomenon only within families of Hamiltonians that support at least two parameters, since it requires $x=y=0$.    The jargon for this is that crossing is a codimension two phenomenon.  If we have an $n$-dimensional parameter space, the expectation is that crossing will occur on an $n-2$ dimensional submanifold.
\item Near the crossing, the energy surface is not analytic, but has a conical singularity.
\end{enumerate}
The third basic property was observed by Herzberg and Longuet-Higgins.   Now we consider not only the energies, but also how the wave-functions depend on $x,y$.   From Equation (\ref{diagonalization}) we have, continuing the upper (positive-energy) eigenfunction continuously from $\phi = 0$:
\begin{eqnarray}\label{psiPlus}
\psi_+ (x, y) ~&=&~ e^{-i\frac{\phi}{2} \sigma_2} \left(\begin{array}{c}1 \\ 0 \end{array}\right) \\
~&=&~ \left(\begin{array}{c}\cos \frac{\phi}{2} \\  \sin \frac{\phi}{2}\end{array}\right)
\end{eqnarray}
Therefore

$ \ \ \ \ $ 3. As $\phi$ evolves from $0$ to $2\pi$  -- as we circle around the crossing -- the sign of $\psi_+(x, y)$ reverses.  

The sign change is a {\it discrete, topological\/}  feature of the quantum geometry.  Thus the crossing leaves a robust signature, that extends far away in parameter space.   This structure is known in other contexts, and has a name: we've found a $Z_2$ vortex. 

This whole set-up relied on restricting to real wave-functions (and real Hamiltonians).   This restriction is appropriate for $T$-invariant systems where spin-dependent interactions are negligible, and no internal symmetries are in play.  If we allow ourselves to use complex wave-functions, it seems that the topology disappears.  We could, specifically, redefine 
\begin{equation}\label{funnyChoice}
\tilde \psi_+ (x, y) ~\equiv~ e^{i\frac{\phi}{2}} \left(\begin{array}{c}\cos \frac{\phi}{2} \\  \sin \frac{\phi}{2}\end{array}\right)
\end{equation}
and have eigenfunctions that both depend smoothly on $\phi$ and remain invariant as $\phi \rightarrow \phi + 2\pi$.   
But by going a little deeper we'll recover and vastly generalize the non-trivial topology suggested by our earlier, ``more natural'' choice.

\subsection{Geometric Matrix}

\subsubsection{A General Formulation}

The geometric matrix or Berry phase is discussed in many places, including the classic \cite{shapere}.   Here I will give a self-contained and slick but rather abstract derivation; if the whole subject is unfamiliar you might want to consult the literature.   

The proper dynamical setting for level-crossing problems is adiabatic evolution.   We consider Hamiltonians with discrete levels, some of which may be multiply degenerate.  (That will be all-important later.)  As the parameters vary, the values of the energies associated to levels will change, as will their position in Hilbert space, but the ordering of levels will not change, except of course at crossings, where the adiabatic approximation breaks down.  Thus we consider
\begin{equation} \label{diagonalizedHamiltonian}
H ~=~ S \Delta S^{-1}
\end{equation}
with
\begin{equation}
\Delta \, = \, \left(\begin{array}{cccc} E_1 & 0 & 0 & ... \\0 & E_2 & 0 & ... \\0 & 0 & E_3 & ... \\... & ... & ... & ...\end{array}\right)
\end{equation}
a {\it block}-diagonal matrix, and $S$ a unitary transformation.   This is a perfectly general form for an Hermitean Hamiltonian, of course.   It is understood that $S$ and $\Delta$ are functions of time.   

Now defining
\begin{equation}
\psi (t) \, =  \, S(t) \left(\begin{array}{cccc} e^{-\frac{i}{\hbar} \int^t_0 E_1(s) ds} & 0 & 0 & ... \\0 &  e^{-\frac{i}{\hbar} \int^t_0 E_2(s) ds} & 0 & ... \\0 & 0 & e^{-\frac{i}{\hbar} \int^t_0 E_3(s) ds}  & ... \\... & ... & ... & ...\end{array}\right)  \tilde \psi (t) 
\end{equation}
or more briefly
\begin{equation}
\psi (t) \, = \, S(t) e^{-\frac{i}{\hbar} \int^t_0 \Delta (s) ds} \tilde \psi (t) 
\end{equation}
and applying the Schr\"odinger equation to the column vector $\psi$, we arrive at 
\begin{equation}\label{neatEquation}
\dot {\tilde \psi} \, = \, - e^{+\frac{i}{\hbar} \int^t_0 \Delta (s) ds} S^{-1} \dot S e^{-\frac{i}{\hbar} \int^t_0 \Delta (s) ds}   \tilde \psi . 
\end{equation}
In the product of matrices on the right-hand side the $ij$ component will contain a factor 
$$
e^ {\frac{i}{\hbar} \int (E_i - E_j) } 
$$
from the $\Delta$ integrals.
If $ S^{-1} \dot S$ is slowly varying, and $E_i - E_j$ is not too close to zero, the contribution of this oscillatory term, which picks out frequency $\frac{E_i - E_j}{\hbar}$, to the evolution will be highly suppressed.  The adiabatic theorem in its crude form -- ``no quantum jumps'' --  instructs us to drop such terms.   That principle does not give us a unique solution, however, even for infinitely slow changes of parameters.  

Defining $\Pi^{( \kappa )}$ to be the projection operators that restrict to the $\kappa$th level of degenerate eigenvalues, the adiabatic approximation in its crude form gives us the approximate equation
\begin{equation} \label{tildeWavefunctionEquation}
{\dot {\tilde \psi}}^{(\kappa)} \, = \, -  \Pi^{(\kappa)} S^{-1} \dot S \Pi^{( \kappa )}  {\tilde \psi}^{(\kappa)}  
\end{equation}
It is useful at this point to introduce the gauge potentials 
\begin{equation}
A^{(\kappa )}_j \, \equiv \, - \Pi^{(\kappa )} S^{-1} \frac {\partial S}{\partial \lambda^j} \Pi^{(\kappa )} , 
\end{equation}
where the $\lambda^j$ parameterize the space of transformations $S$.   
In terms of the $A^{(\kappa )}_j$ we can solve \eref{tildeWavefunctionEquation} in terms of a path-ordered integral
\begin{equation}
(\tilde \psi )^{( \kappa )} (t) \, ~=~ \, P \left[\exp \int_{S(0)}^{S(t)} d\lambda^j A^{(\kappa )}_j\right] (\tilde \psi )^{( \kappa )} (0), 
\end{equation}
and therefore
\begin{equation}\label{finalWavefunctionSolution}
\psi ^{( \kappa )} (t) \, ~=~ S(t) e^{-\frac{i}{\hbar} \int^t_0 \Delta (s) ds} \, P\left[ \exp \int_{S(0)}^{S(t)} d\lambda^j A^{(\kappa )}_j\right] S^{-1}(0) \psi^{( \kappa )} (0) .
\end{equation}

The characteristic factor 
\begin{equation}
{\cal C}({\rm path}) ~\equiv~ P \left[\exp \int_{S(0)}^{S(t)} d\lambda^j A^{(\kappa )}_j\right]
\end{equation}
connects the eigenspaces for energy $E_\kappa$, as they vary with $\lambda$, in a fashion reminiscent of Wilson lines in gauge theory or parallel transport in Riemannian geometry.   In the present context, this connection is known as the ``Berry phase'' or ``geometric phase'' or ``geometric matrix'' (which I'll use).   ${\cal C}$ has several remarkable properties: 
\begin{itemize}
\item It depends only on the geometry of the embeddings, and not on how fast they are run through.
\item It remains non-trivial no matter how slow the adiabatic evolution is.
\item It is of order $\hbar^0$.   The energy$\rightarrow$frequency factors go as $\hbar^{-1}$ (inside the exponential).   Higher corrections, with positive powers of $\hbar$, are non-adiabatic, in the sense that they can become arbitrarily small as the evolution is taken arbitrarily slow.
\end{itemize}

Note that in Equation (\ref{finalWavefunctionSolution}) the geometric matrix operates inside the matrix $S$.  This shows that it is more properly considered as a correction to the basis-geometry than as a correction to the dynamics.  

\subsubsection{Back to the Vortex}

Returning to our example: If we use the wave functions from \eref{funnyChoice} then we find that the geometric matrix factor, integrated from $\phi = 0$ to $2\pi$, restores the minus sign that the original, totally real analysis suggested.   The calculation is simple:
\begin{eqnarray}
A_\phi ~&=&~ 
- \left(\begin{array}{cc}\cos \frac{\phi}{2}  & \sin \frac{\phi}{2} \end{array}\right) e^{-i \frac{\phi}{2}} \frac{\partial}{\partial \phi} e^{i \frac{\phi}{2}}\left(\begin{array}{c}\cos \frac{\phi}{2}  \\\sin \frac{\phi}{2}\end{array}\right) ~=~  -\frac{i}{2} \nonumber \\
e^{\int d\phi A_\phi} ~&=&~  e^{-i\frac{\phi}{2}}
\end{eqnarray}
The field strength associated to this gauge potential vanishes; nevertheless it is globally non-trivial, like the electromagnetic vector potential in the Aharonov-Bohm effect.

On the other hand if we stick to the real wave functions, then we find that we cannot make a smooth choice consistently over the whole circle.  We must introduce a change in sign somewhere.   The usual procedure is to cover the circle with two overlapping patches, say over $-\pi - \delta < \phi < \delta$ and $-\delta < \phi < \pi + \delta$, with $\delta$ a small positive number.   On each patch we can use the $\psi_+$ of Equation (\ref{psiPlus}).   In this formulation, the geometric matrix is just the identity, but there is a nontrivial transition-factor at the overlap, since the wave functions on the two patches differ by a sign around $\phi = \pm \pi$.

So we have a choice.  We can use a global, smooth basis for the wave functions, together with a non-trivial geometric matrix; or alternatively, a smooth real-valued basis, defined in patches, with no geometric matrix but with a non-trivial transition on the overlap.  They are completely equivalent descriptions, as our derivation of both from the refined adiabatic theorem demonstrates.  (The second formulation illustrates how we construct gauge theories based on discrete gauge groups.)   Both formulations lead to the conclusion that we have a $Z_2$ vortex surrounding crossings of non-degenerate levels governed by real Hamiltonians.

\subsection{The General Case}

The crossing problem is only non-trivial for states of the same symmetry.  But given that restriction, we still have many possibilities beyond the original case considered by von Neumann and Wigner.   Indeed we can consider different symmetry groups and representations, and crossing problems for multiplets in each class.    (Here I'm summarizing some results from a forthcoming paper with Kam Luen Law \cite{LawFW}.  Earlier, related work includes \cite{simon} \cite{aitchison}.) 

At first glance one might expect that the geometry could get very complicated, especially for multiplets of large dimensions, and that there are many cases to consider.   It's pleasant that closer analysis reveals that the complexity is relatively modest, and that there is a remarkable degree of universality in the geometry governing different multiplet situations.  (For sophisticates: The mathematical fact that keeps things under control is Schur's lemma; and the key to proving the statements that follow is character theory.)   This is not the place to enter into technicalities or proof; I will just indicate the general result and exemplify one interesting case.  


\begin{table}[htdp]
\caption{\label{table 1} Group representations, assumed unitary and irreducible, can be classified as real, pseudoreal (alternatively: quaternionic), or complex.   Real representations, of course, are representations that can be defined using only real numbers, such as the vector representation of $SO(3)$.  Such representations are, of course, isomorphic to their complex conjugates.    Pseudoreal representations are representations that are isomorphic to their complex conjugates, but which cannot be implemented using only real numbers.   The spinor representation of $SU(2)$ is a familiar example.   Representations that are not isomorphic to their complex conjugates, like the {\bf 3} of $SU(3)$, are said to be complex.   (Another characterization: A representation is real if its symmetric tensor product with itself contains the identity; pseudoreal if its antisymmetric tensor product with itself contains the identity; and complex otherwise.)  Time reversal symmetry $T$ is implemented by an antiunitary transformation, and one may have $T^2 = \pm 1$.  Depending on the complexity-type of a representation, and the sign of $T^2$, it may or may not be necessary to double the representation in order to implement $T$ (generalized Kramers theorem).  Hamiltonians invariant under the group and under $T$ can connect two multiplets of the same type only in restricted ways, that can be described by a finite number of parameters.    We find that in different cases the geometric matrix field describes a vortex, monopole, or instanton, as indicated.}
\begin{indented}
\item[]\begin{tabular}{ccccc}
\br
multiplet type & $T^2$ & doubled? & parameters to fix &  topology\\
\mr
real & 1 & no & 1 + 1& $Z_2$ vortex\\
\mr
real & -1 & yes & $4+1$ & $SU(2)$ instanton\\
\mr
complex & 1 & yes & $2+1$ & $U(1)$ monopole\\
\mr
complex & -1 & yes & $2+1$ & $U(1)$ monopole\\
\mr
pseudoreal (quaternionic) & 1 & yes & $4+1$ & $SU(2)$ instanton\\
\mr
pseudoreal (quaternionic) & -1 & no & $1+1$ & $Z_2$ vortex\\
\br
\end{tabular}
\end{indented}
\end{table}%

The general situation is indicated in Table 1 and its caption.   Here I'll just spell out the second line a bit, for the special case of the identity representation.   

If we want to implement $T^2 = -1$, we can't use a one-dimensional space, since $| \psi \rangle$ and $T | \psi \rangle$ are orthogonal:
\begin{equation}
\langle \psi | \, (T | \psi \rangle ) ~=~ - ( \langle \psi | T^{\dagger 2} ) \,  (T | \psi \rangle ) ~=~ - \bigl( ( \langle \psi | T^\dagger) T^\dagger \bigr)\,  (T | \psi \rangle ) ~=~  - \langle \psi | \, (T | \psi \rangle )
\end{equation}
where in the last equation we've used the antiunitarity of $T$, viz. $( \langle \alpha | T^\dagger)(T | \beta \rangle) = \langle \beta | \alpha \rangle$.  For a minimal implementation we can take $T = K\,  i\tau_2 $ within a two-dimensional space, where $K$ denotes complex conjugation.   (A physical motivation for $T^2 = -1$ is a sort of converse to this construction: If we want $T$ to reverse all three spin operators $\sigma_j$, and yet to include complex conjugation, we must have $T = K \, i \tau_2 $ acting on the spin variable, so that if the rest of $T$ is ``normal'' we'll have $T^2 = -1$.)   

We have two-complex dimensional multiplets, and we want to consider crossing, so we have two doublets.  Altogether then we've got a four-dimensional Hilbert space, with $T$ realized as 
\begin{equation}
T ~=~ K i\tau_2 \otimes 1
\end{equation}
Now we can consider the space of possible Hamiltonians, i.e. Hermitean operators, invariant under $T$.  Besides the identity, we easily identify five others:
\begin{eqnarray}
\Gamma_1 ~&=&~ \tau_1 \otimes \tau_2 \nonumber \\
\Gamma_2 ~&=&~ \tau_2 \otimes \tau_2 \nonumber \\
\Gamma_3 ~&=&~ \tau_3 \otimes \tau_2 \nonumber \\
\Gamma_4 ~&=&~ 1\ \otimes \tau_1 \nonumber \\
\Gamma_5 ~&=&~ 1\ \otimes \tau_3
\end{eqnarray}
and one can prove that these form a basis for all invariant Hamiltonians.    

As the notation is meant to suggest, there's a pretty surprise here: These Hamiltonians obey the five-dimensional Clifford algebra
\begin{equation}
\{ \Gamma_j, \Gamma_k \} ~=~ 2 \delta_{jk}
\end{equation}
Thus we see that the  space of relevant Hamiltonians near a crossing is the same as in Equation (\ref{sigmaForm})
\begin{equation}
H(x^j) ~=~ x^j \cdot \Gamma_j
\end{equation}
but now on a five-dimensional rather than a two-dimensional space.   

In parallel with our earlier discussion, we can read off two basic conclusions:  The crossing is {\it highly\/} avoided -- codimension five! -- and the energy function has a conical singularity:
\begin{equation}
E(x^j) ~=~ \pm \sqrt { \vec x \cdot \vec x } 
\end{equation}

The most interesting issue, intellectually, is the geometric matrix.   Since the eigenfunctions (as opposed to the eigenvalues) do not depend on the magnitude of $\vec x$, and we excise $\vec x = 0$, the relevant parameter space is the sphere $S^4$.   The positive and negative eigenspaces are each two-dimensional, so we will get a gauge potential in the group $U(2)$.   Because the underlying Clifford algebra is $SO(5)$ invariant, so will be the gauge structure we derive from it.    So we can anticipate that a very special gauge structure will emerge.   

Detailed calculation confirms that anticipation.  Parallel transport, in the sense of Riemannian geometry, gives us an ($SO(5)$ invariant) $SO(4)$ gauge structure on $S^4$.   Using the Lie algebra isomorphism $SO(4) \rightarrow SU(2) \times SU(2)$, and projecting on either factor, we get an ($SO(5)$ invariant) $SU(2)$ gauge potentials.  The geometric matrices  in the positive and negative eigenspaces are governed by precisely those two $SU(2)$ potentials.   The resulting gauge structures are the (sphere versions of) instantons and anti-instantons, famous in quantum field theory and topology.    Because these gauge structures are topologically non-trivial, they leave robust signatures even far from the crossing.

\subsection{Comments}

\begin{enumerate}
\item It might seem extravagant to consider structures that are so hard to get at, i.e. with such high codimension.   Where are all those parameters supposed to come from?   Well, they could be the positions of quasiparticles (as occurs, especially, in anyon physics), quasimomenta in a Brillouin zone, external electric or magnetic fields, or even ``conceptual parameters'' that allow us to interpolate, theoretically, between materials with fundamentally different Hamiltonians (this is an important idea in the theory of topological insulators), ...    
\item One can often relate the occurrence of level crossings, as a function of pumping parameters, to transport.  In that context, topological invariants associated to the level crossing can govern the discrete parameters that appear in quantized transport.  
\item A common source of level {\it merger} -- a more general version of crossing -- is the occurrence of enhanced symmetry.  For example, in the context of band theory one often finds enhanced symmetry at special points in the Brillouin zone.   If the symmetry of the Hamiltonian, as a function of parameters $\lambda$, is enhanced from $H$ to $G$ at $\lambda_0$, then irreducible representations of $G$ will be formed by merging $H$-multiplets that are non-degenerate away from $\lambda_0$.  Interesting geometric matrices can arise this way -- and in graphene, specifically, they do.
\end{enumerate}


\begin{thebibliography}{99}

\bibitem{thomsonTait} Epple M 1998 {\it Arch. Hist.  Exact Sci}. {\bf 52}  297

\bibitem{GW} Goldstone J, Wilczek F 1981 {\it Phys. Rev. Lett}. {\bf 47} 986

\bibitem{witten} Witten E 1979 {\it Phys. Lett. B} {\bf 86} 283


\bibitem{jr} Jackiw R, Rebbi C 1976 {\it Phys. Rev. D} {\bf 13} 3398

\bibitem{vNWigner}  von Neumann J, Wigner E 1929 {\it Phys. Zeit}. {\bf 30} 467

\bibitem{HerzbergL-H} Herzberg G, Longuet-Higgins H C 1963 {\it Disc. Faraday Soc}. {\bf 35} 77


\bibitem{shapere} Shapere A, Wilczek F 1989 {\it Geometric Phases in Physics\/} (World Scientific)

\bibitem{LawFW} Law K T, Wilczek F {\it paper in preparation}

\bibitem{simon} Avron J, Sadun L, Segert J, Simon B 1989 {\it Comm. Math. Phys}. {\bf 124} 595
 
\bibitem{aitchison} Johnsson M, Aitchison I 1997 {\it J. Phys. A} {\bf 30} 2085




\end{thebibliography}
\end{document}